\documentclass[12pt]{iopart}
\usepackage{iopams}  
\usepackage{graphicx}
\usepackage[dvipsnames]{xcolor}

\begin{document}
	
	
	\title{An eccentric binary blackhole in post-Newtonian theory.}
\author{Sourav Roy Chowdhury}

\address{Research Institute of Physics, Southern Federal University, 344090 Rostov on Don, Russia.}
\ead{roic@sfedu.ru}
	
\author{Maxim Khlopov}
\ead{khlopov@apc.in2p3.fr}
\address{Research Institute of Physics, Southern Federal University, 344090 Rostov on Don, Russia.\\ APC Laboratory 10, rue Alice Domon et L\'eonie Duquet, 75205 Paris Cedex 13, France. \\Center for Cosmoaprticle Physics Cosmion, Moscow State Engineering Physics Institute, National ResearchNuclear University “MEPHI”, 31 Kashirskoe Chaussee, 115409 Moscow, Russia}

	\date{\today}
	
	\begin{abstract} Gravitational waves radiated during binary black hole coalescence is a perfect probe for studying the characteristics of strong gravity. Advanced techniques for creating numerical relativity substitute models for eccentric binary black hole systems are presumed to become more crucial in existing and anticipated gravitational wave detectors. The imprint on the observation data of the gravitational wave emitted by the binary coalescence enhances the two body system studies.  The aim of this study is to present an overview of the change in characteristics behaviours of hierarchical massive astrophysical objects merger, which are the data bank of early universe. We present results from numerical relativity simulations of equal-mass and unequal mass nonspinning inspiral binary-black-hole system in the Post Newtonian framework. We also consider the time evolution of eccentricity for initial eccentric system. The eccentric Post Newtonian equations are expanded in the form of frequency related variable $ x = (M \omega)^{2/3}$. The model is restricted to the (2, 2) spin-weighted spherical harmonic modes. We conclude that for higher eccentricity as well as mass ratio there exist higher oscillation in orbital radius and in eccentricity.
	\end{abstract}

	\maketitle
	
	
	\section{Introduction}	
	
	Detection of gravitational wave (GW) signals with the Advanced LIGO and VIRGO, and KARGA detectors   \cite{abbott9,abbott8,abbott7,Ligo1,abbott1,abbott2,abbott3,Abadie} provides the prospect of confirming gravity theories. The detectors are expected to detect a network of  coalescing comparable mass compact binaries (black hole--black hole mergers, neutron star--neutron star mergers, or black hole--neutron star mergers) \cite{abbott4,abbott5,abbott6}.
	
	The stochastic gravitational wave background (SGWB) is expected to be produced by the contributions of superposition from several distinct and unresolved GW sources. The SGWB is a very well-known cosmological framework. It appears in various inflationary \cite{Grishchuk,Starobinskii,Barnaby} and cosmic string models \cite{Caldwell,Damour1,Vuk}. It could be astrophysical, created by a number of of astrophysical sources, such as compact binary coalescences (CBCs)\cite{Regimbau2,Regimbau3,Rosado2,Vuk1}; it may also be from a neutron star \cite{Vuk2,Rezzolla,Rezzolla1,Cheng,chowdhury} or from early instabilities \cite{Owen,Howell,Ferrari}.
	
	The signature of the eccentricities is not very typical in the observed detections. Different researchers have shown non-negligible eccentric signatures in the compact binaries formed in sufficiently dense stellar clusters \cite{Rodriguez}. 
	The eccentricities have even been restricted in some studies on observed events \cite{Gayathri,Nitz,Romero,Romero1}. It has been found that roughly half of all Binary Black holes (BBHs) mergers happen within the cluster, with around 10\% of such mergers crossing the LIGO/Virgo band with eccentricities more significant than 0.1. The result is based on modeling relativistic accelerations and GW emission for isolated binaries and three- and four-body interactions. Eccentric binaries are potential sources of GW in future observational cycles, which can retain the signature of their creation channels. As the leftovers from BBHs mergers, black holes with masses inside this ``high mass gap'' are projected to develop \cite{Downing,Downing1,Rodriguez1,Banerjee}. For heavily populated situations such as globular clusters, the remnants can dynamically merge again \cite{Belczynski,Mandel,Mandel1}. The considerable mass of the BBH merger GW190521 supports the theory of evolving dynamically. A binary that merges quickly and then becomes bonded may not circularize before merging. Orbital eccentricity can also suggest dynamic production.

	The inspiral, merger, and ringdown phases are represented by three waves in the waveform of a merging BBH. The mass information is encoded in the GW frequency ($ f $) and its time derivative ($ \dot{f} $) during the inspiral phase. EccentricTD \cite{Tanay}, a time-domain, inspiral-only, and nonspinning waveform model, valid for high eccentricities and second-order post-Newtonian corrections, is one of several eccentric waveforms. EccentricFD \cite{Huerta} is an analytic, frequency-domain, inspiral-only, and nonspinning model that can handle eccentricities up to 0.4. A few other numerical relativity waveforms are practical for numerical values of eccentricity, spin, mass ratio, etc. \cite{Boyle}. The models seem unsatisfactory for eccentricities and spins. The state-of-the-art literature restricts the scope of detailed parametrization for a wide range of values. For different models of merging Population III BBHs, the chirp mass ($ M_{chirp} = (M_1M_2)^{3/5}/(M_1 +M_2)^{1/5}$) distributions are almost the same by nature. The maximum chirp mass of merging of Population  III BBHs is around 30 M$_\odot $. This remains unaffected by the primary conditions of the evolution variables of the binary system \cite{Kinugawa}. The features of Population III's dynamic passage cause this tendency. Normalized intrinsic chirp mass distributions are also maximized for chemically homogeneous evolution models~\cite{Ossowski}. Bose and Pai \cite{pai} introduced a model-independent GW searches chirp-cut, which is important for discriminating  compact binaries of inequivalent mass ranges. The chirp mass is estimated from the time-frequency representation. The model can confine eccentricity by using quasicircular nonspinning waveforms.

	The growth of linear perturbations on the spacetime metric of the remnant is referred to as ringdown \cite{Pani}. The dynamics in the strong-field regime that established the perturbation constrains the ringdown. 
	The properties of the remnant BH are not independent of an underlying theory of gravity. General relativity is changed in the highly nonlinear framework. In this case, the relative amplification of modes in ringdown, the mass, and the spin of the final BH should change \cite{Kamaretsos,Hughes,Detweiler,Kamaretsos1}. Bhagwat and Pacilio \cite{Bhagwat} proposed a test to verify the perturbation conditions during the plunge-merger phase match. The BH properties remained after the ringdown phase. The proposed evaluation confirmed the consistency by examining the frequency response as well as the amplitudes and phases of amplification in the ringdown signals simultaneously.

	In order to evaluate the source parameters from GW data by using the matched filtering method, meticulous models of the waveforms are required. Models developed based on the post-Newtonian (PN) approach are used for inspiral eccentric binaries \cite{Arun1,Arun2,Arun3,Hinder,Gopakumar,Tessmer,Tanay,Hinder1}. Binaries in eccentric orbits have a well developed PN theory. On the other hand, PN can only simulate the inspiral waveform, since the assumption considers that the black holes are spaced and gradually drifting. 
	
	The detectors require a more precise model to match the observed events of GW from merging black holes. Only numerical simulations of  Einstein's equations of general relativity can quantify the dynamics of black hole mergers and eventually the GW form. Huerta et~al. \cite{Huerta1} introduced the eccentric waveform model combining inspiral, merger, and ringdown. It has been assumed persistently that the eccentricity will become negligible by the time of merging. The  \textit{inspiral-merger-ringdown} 
	(IMR) model is alternatively called the \textit{``ax-model''}. However, work is still ongoing to develop complete IMR waveforms, along with spins and precession to the maximum eccentricity. If there exists a considerable eccentricity at the final stage of the inspiral of the BBH system, it cannot be measured precisely. A circular binary black hole inspiral creates waveform dynamics. The adiabatic estimation performs well because of the monotonically increasing behavior of amplitude and frequency. Eccentric orbits on either side produce waveforms with oscillating amplitude and frequency. Our eccentric model is an inspiral waveform model evaluated for different mass ratios, calibrated to fit the time between the waveform approaching a given frequency and the maximum waveform amplitude in the transition region of the eccentric PN portion. The inspiral is established using the PN, with a superior version of the \textit{``x-model''} \cite{Hinder}. The noneccentric Implicit Rotating Source model has been used to model the merger \cite{Kelly}. 
	
	Due to its richness in intriguing theoretical aspects, the gravitational two-body problem extends beyond its applications to GW physics. The configuration of hierarchical mergers that develop more massive ones is one of the dynamical scenario's distinctive signatures. Only dense star clusters allow this process, as the merger remnant, previously a single BH, can obtain a companion through dynamical exchanges. The characteristic changes in behavior are, thus, also important to understand.
	
	The paper is organized as follows. The basic structure of the IMR model is developed in Section \ref{sec2}. Once the main features are identified, we describe the PN x-model. Direct comparison with three PN waveforms implies the rationality for using different mass ratios and eccentricities. The comparison of the separation of BBH systems in an effective one-body framework for eccentric binary PN waveforms is presented in Section \ref{sec3}. Finally, we conclude  in Section \ref{sec4}.
	
	The natural unit system has been used throughout in the work.
	
	\section{Basic Structure of the PN Model} \label{sec2}
	
	We start  to discuss the eccentric PN inspiral model from eccentric Newtonian orbits. Let us  consider a two-body system that consists of masses $m_1$ and $m_2$ at the points $\vec{x_1}$ and $\vec{x_2}$. The symmetric mass of the system can be defined as $\eta = \mu/M $, where $\mu = m_1 m_2/M$, the reduced mass of the system, and $ M = (m_1+m_2) $ is the total mass. The mass ratio ($m_1/m_2$) is defined as  $ q $. The separation between the points ($ r = |\vec{x_1}-\vec{x_2}| $)  satisfies the equation of an orbital radius,	
	\begin{equation}
	r = a (1-e \cos u).\label{e1}
	\end{equation}
	
	The angular variable $ u $ is the eccentric anomaly, which characterizes the oscillatory phase. The pericenter and the apocenter are represented by $ u = 0 $ and $ u = \pi $, respectively. The semi-major axis is defined by $ a $, and $ e $ is the eccentricity. 
	The angular velocity of the system,	
	\begin{equation}
	\dot{\phi} = \frac{n \sqrt{1-e^2}}{(1-e \cos u)^2},\label{e2}
	\end{equation} 
	
	\noindent where `$n$' is the mean motion of the system, is related to the orbital period $ (P) $ by $n=2\pi/P=a^{-3/2}M^{1/2}$.
	
	For the Newtonian system, the angular momentum $ L $ and energy $ E $ are constants of motion. The system is bounded to orbit in the x-y plane (here, nonspinning) as $ L $ is conserved.
	
	The mean anomaly $ l $ of the eccentric orbit can be defined in terms of $ u $ from Kepler's equations as follows,	
	\begin{equation}
	l = 2 \pi \Big(\frac{t - t_0}{P}\Big) = u - e \sin u,\label{e3}
	\end{equation}
	where $ t_0 $ is the time corresponding to the pericenter passage. Once $ u $ is fixed, solving numerically for time $ t= t_0 $, the $ \dot{\phi} $ and $ r $ are obtained. The constants $ e,~a,~l(t_0),\textrm{~and~} \phi(t_0)$ parameterize each orbit. 
	
	In the PN prescription, the eccentricity is categorized into three types, $ e_t,~ e_r, \textrm{~and}~ e_\phi  $ for the circular motion in $ t, r, \textrm{~and~} \phi $ directions, respectively. (As all three are correlated, setting all three eccentricities equal ($ e \equiv e_i $) does not necessarily deviate substantially from generality. For our purpose, it is sufficient to calculate  $ e_t $ only ). The azimuthal coordinate $ \phi $ is increased by ($ 2 \pi + \Delta \phi $) in each period $ P $, and the pericenter is increased by $ \Delta \phi $. Thus, the average angular velocity can be introduced through $ \omega \equiv (2 \pi + \Delta \phi )/P $. In the circular limit, the angular velocity $ \dot{\phi} $ is not reduced by $ n $. In the relativistic limit for the circular case, $ \omega = \dot{\phi} $; whereas for the Newtonian frame, $ \omega = n $, and $ \omega = constant $, if the radiation impacts are neglected. The expansion of equations in $ x = (M \omega)^{2/3} $ has a good agreement \cite{Hinder}.
	
	Due to the relativistic orbits, the angular velocity relation and Kepler's equation need further modifications in the PN model. As in the Newtonian case, the angular momentum and energy are no longer constant. The gravitational radiation also incorporates the finite changes in the orbits, differing by pericenter precession.  
	
	To calculate the orbit of a binary system that loses both energy and angular momentum, we need to solve a pair of coupled ordinary differential equations (ODEs) in $ x $ and $ e $ . Accordingly, from $ u $, using the PN Kepler equation, $ l,~ r,\textrm{~and~} \phi $ can be obtained. The relevant equations for the PN model are provided in  {Appendix} \ref{appendix}.
	
	To calculate the two distinct gravitational-wave polarization states $ h_{+} $ and $ h_{\times} $, the orientation and the direction of the orbit must be appointed conveniently. We use the GW restricted to the regime $ l=2,~ m=2 $ mode.
	
	The waveform strain for the complex PN system is as follows,
	\begin{equation}
	h = h_+ -ih_{\times}.
	\end{equation} 
	
	The strain $ h $ also decomposes into the spin-weight s = $-$2 spherical harmonic as follows,
	\begin{eqnarray}
	h^{22} = & & \int ~_{-2}Y^{2*}_2(\theta, \varphi) h(\theta, \varphi) d\Omega\nonumber \\
	= & & - \frac{4 M \eta e^{-2 i \phi}}{R} \sqrt{\frac{\pi}{5}} \Big(\frac{M}{r} + (\dot{\phi} r +i \dot{r})^2\Big),
	\end{eqnarray}
	where $ ~_{-2}Y^{2}_2(\theta, \varphi) = \frac{1}{2} e^{-2 i \phi} \sqrt{\frac{5}{\pi}} \cos ^4 (\theta/2) $. $ \theta $ and $ \varphi $  are the spherical polar angles of the observer.

	The system is framed in the $ (t-t_{peak})/M $ coordinate scale.

	\section{Results}\label{sec3}
	
	In this section, we present the results of our analytical study together with the simulation results. The PN model provides limiting waveforms at large distances from the source point. In this study, we focused on classifying the natures based on the mass ratio $q$ and the eccentricity. The simulations were executed for the apocenter. By the comparative measurement of the relevant availible GW strain data  \cite{url}, the model parameters were committed such that the plots reproduced the best data fit for the GW strain. The variables that characterized the system were PN parameters.

\begin{figure}\centering	
		\includegraphics[scale=0.48]{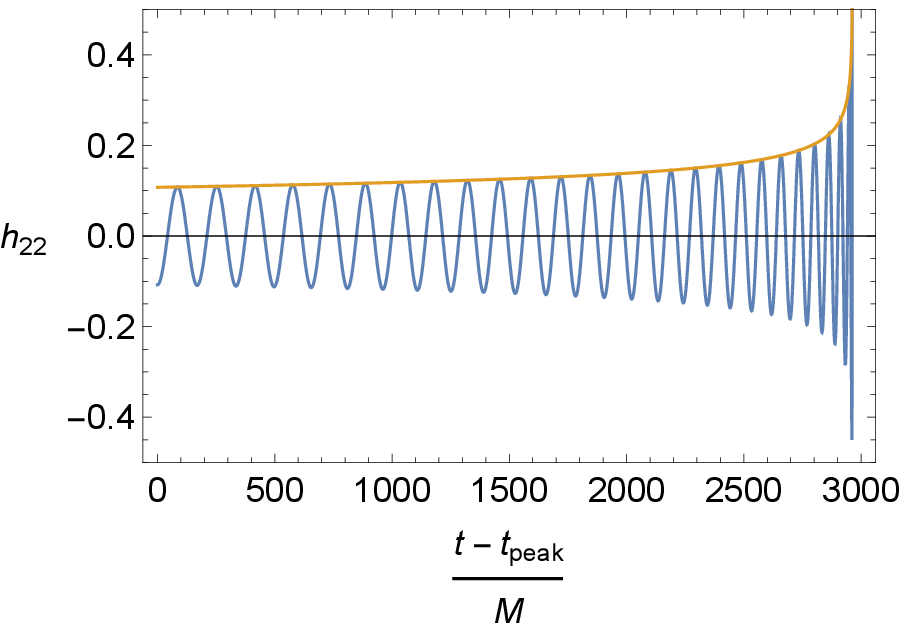}
		\includegraphics[scale=0.48]{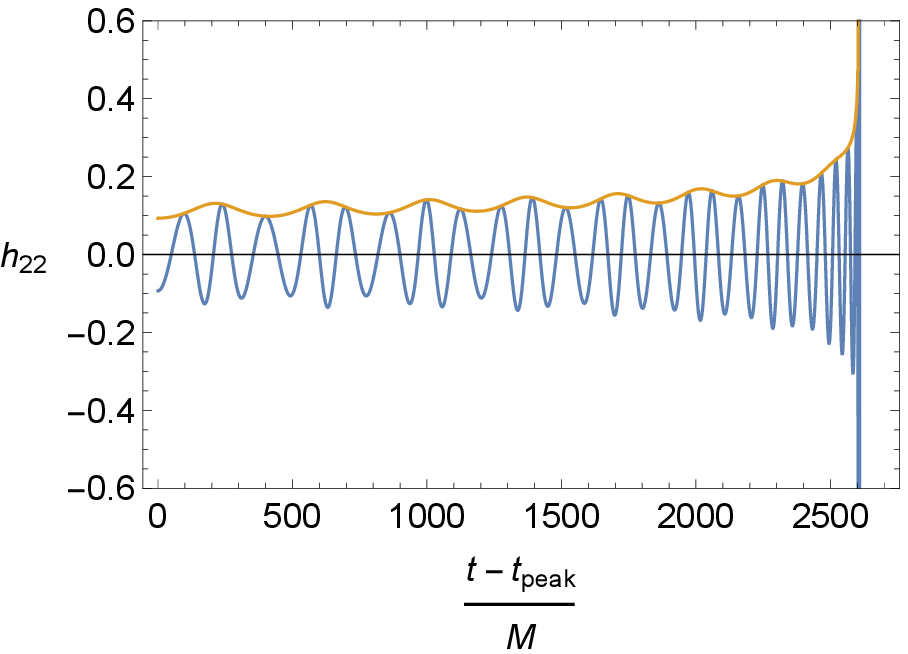}
		\includegraphics[scale=0.48]{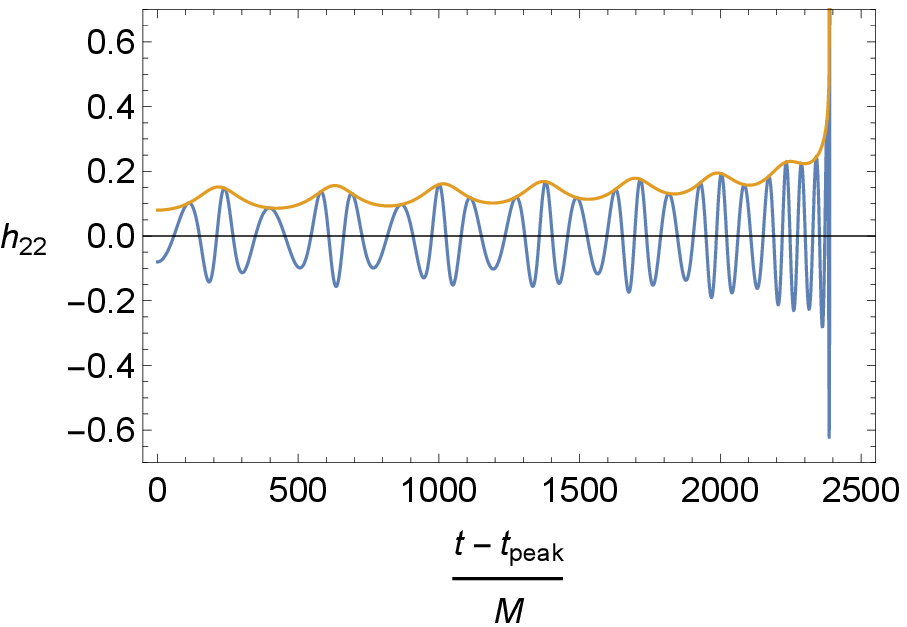}\\
		\includegraphics[scale=0.48]{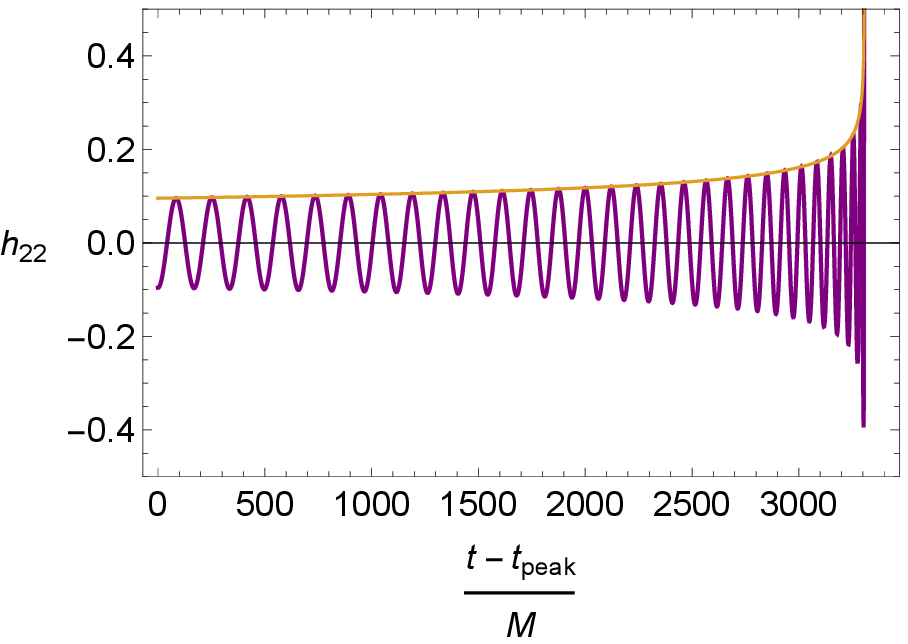}
		\includegraphics[scale=0.48]{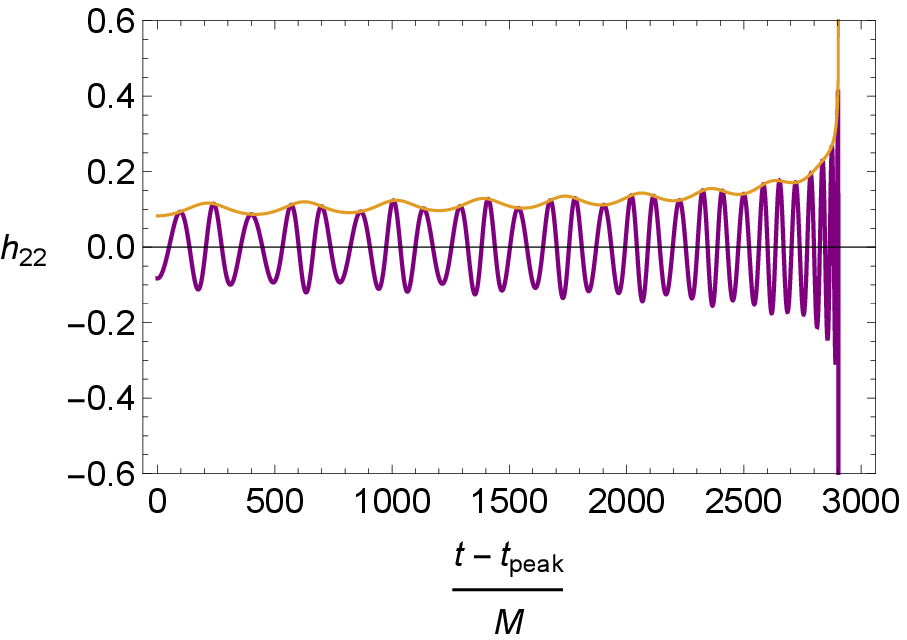}
		\includegraphics[scale=0.48]{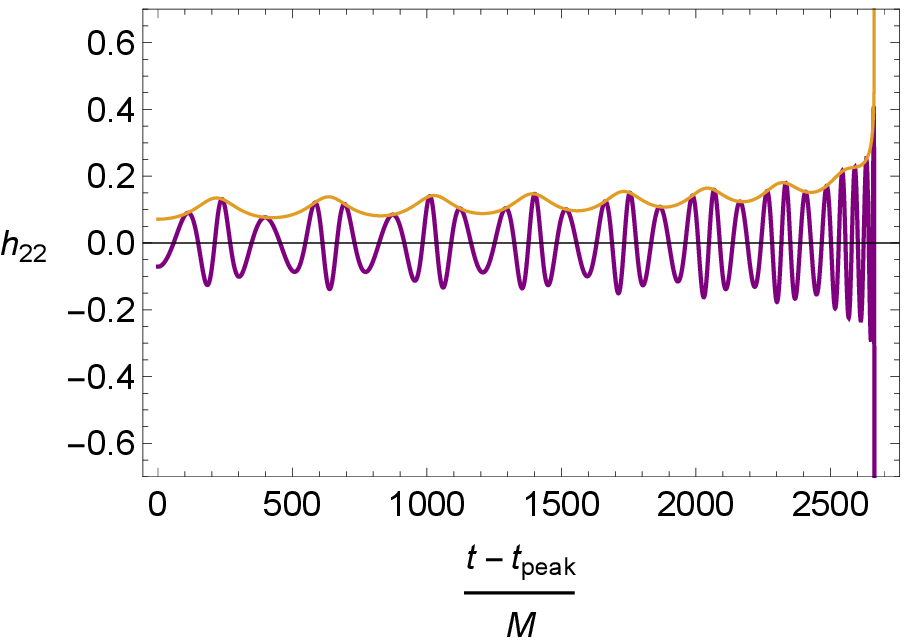}\\
		\includegraphics[scale=0.48]{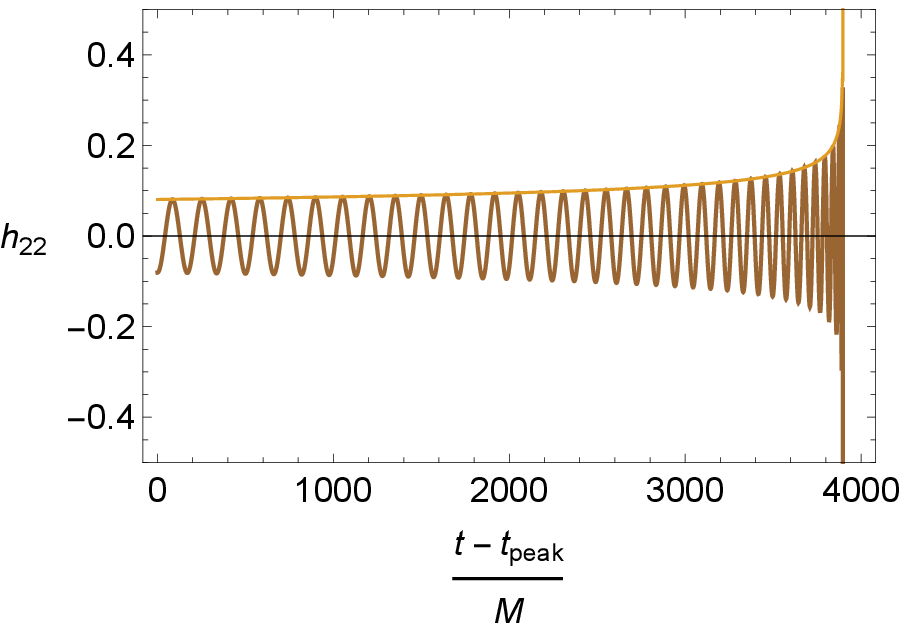}
		\includegraphics[scale=0.48]{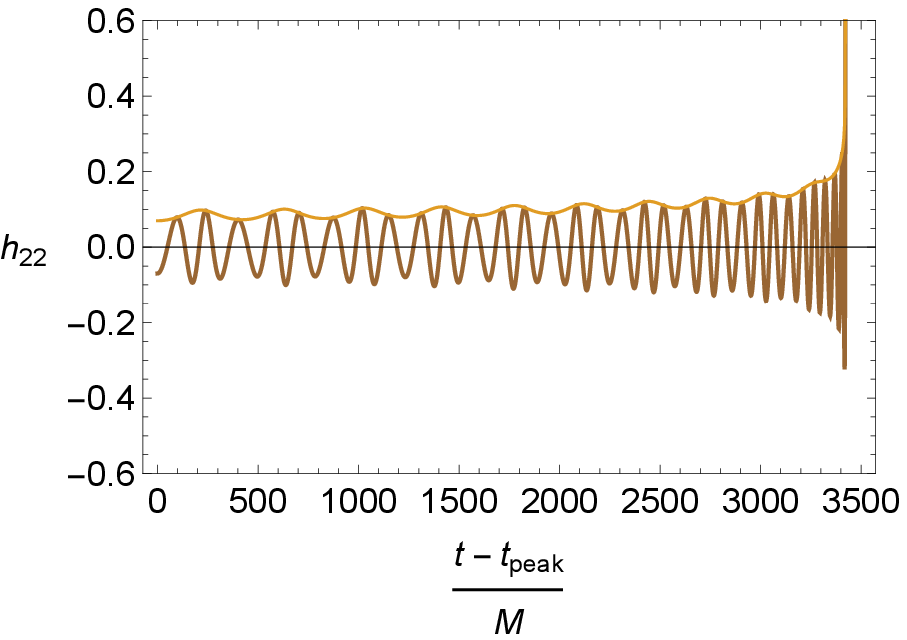}
		\includegraphics[scale=0.48]{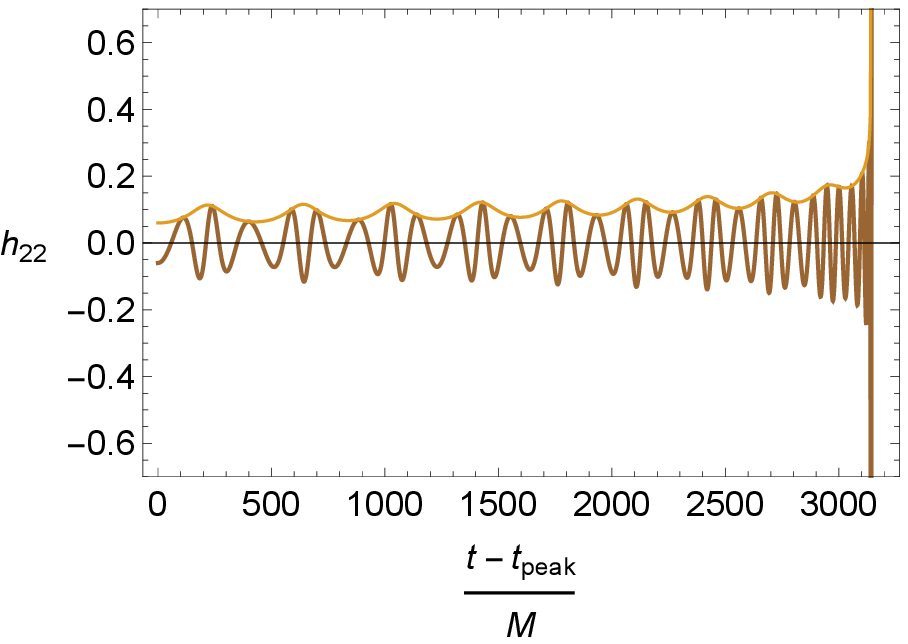}
		\caption{Blue represents $q =1$, purple represents $q = 2$, and brown represents $q = 3$. The left column plots are for $e_{ref} = 0.000$, the middle column plots are for $e_{ref} = 0.100$, and the right column plots are for $e_{ref} = 0.189$. Hereafter, we follow this prescription. The time variation of the GW strain is shown. \label{fh}}
	\end{figure}

	The simulation results for the real part of the strain, $ Re[h_{22}] $ are presented in Figure \ref{fh} for different mass ratios (indicated by different colors). The variations of $ |h_{22}| $ for the same wave strain are also presented in the same figure in saffron.
	
\begin{figure}\centering
		\includegraphics[scale=0.49]{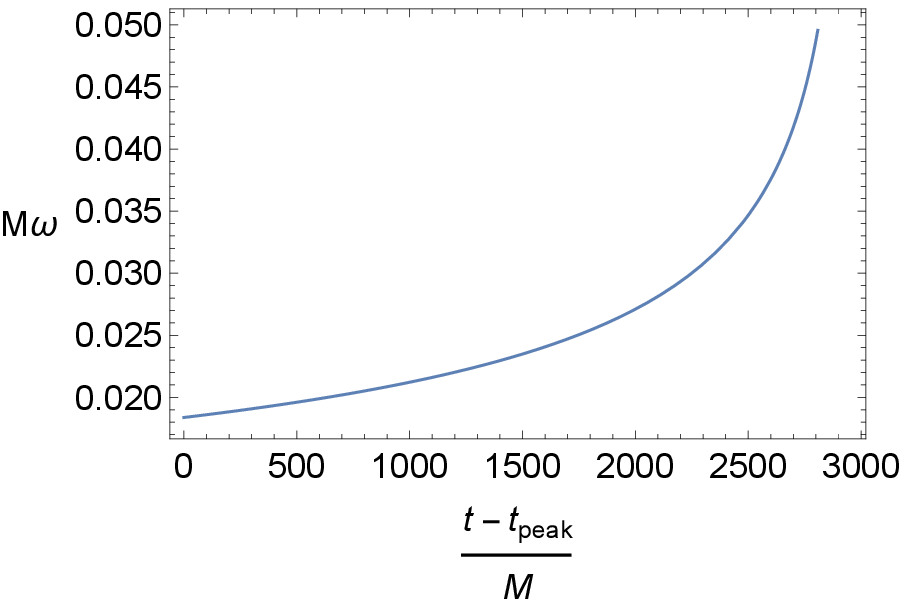}
		\includegraphics[scale=0.48]{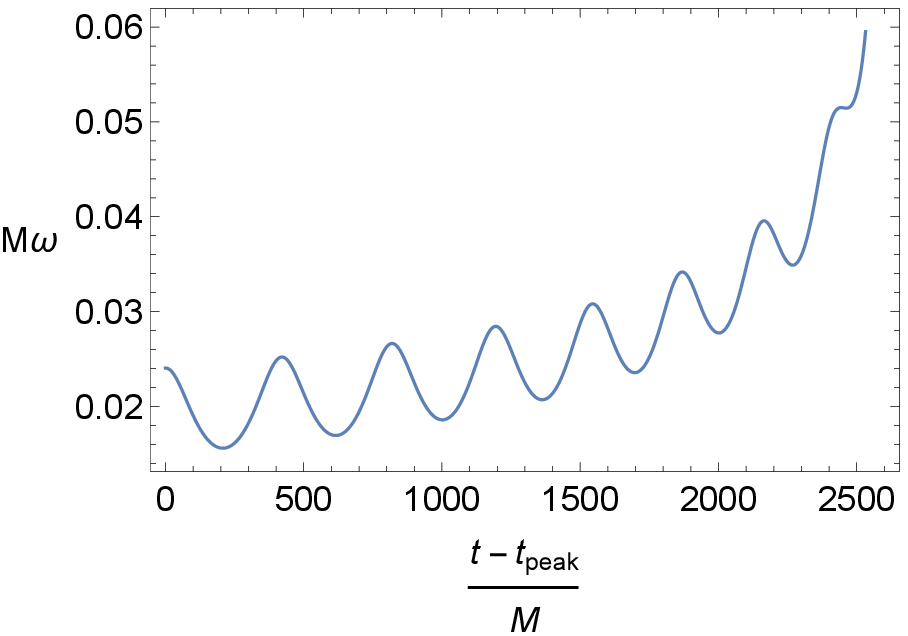}
		\includegraphics[scale=0.48]{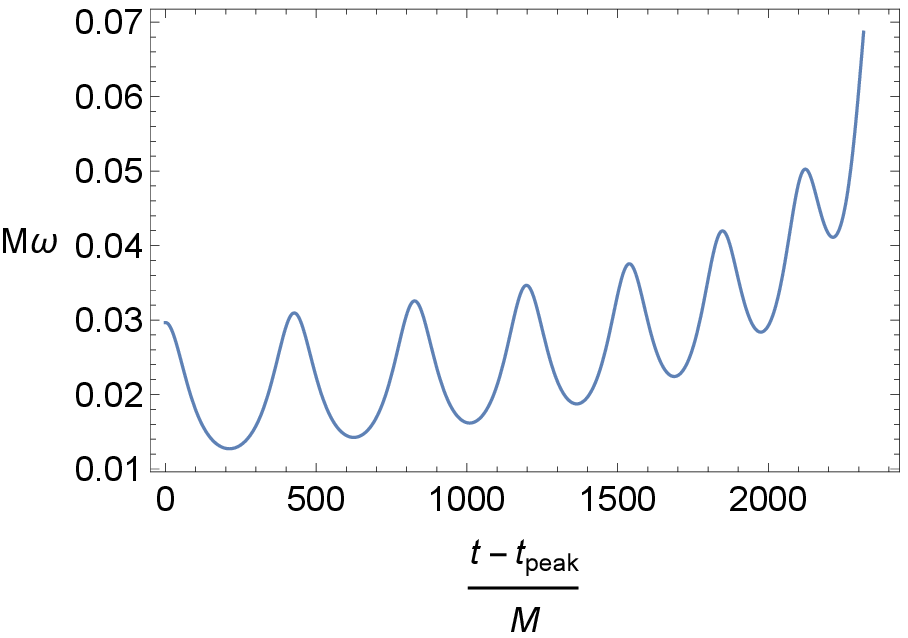}\\
		\includegraphics[scale=0.48]{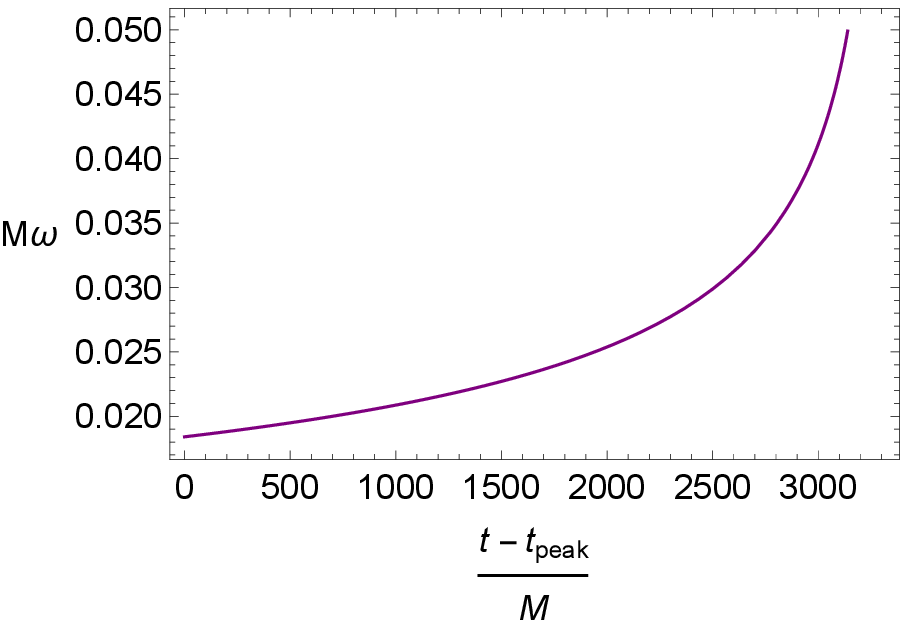}
		\includegraphics[scale=0.48]{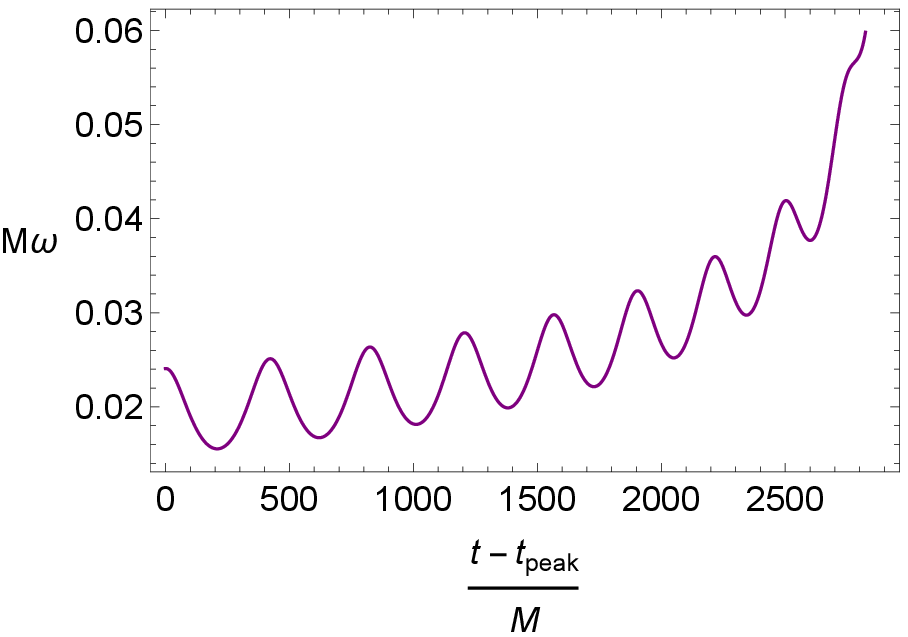}
		\includegraphics[scale=0.48]{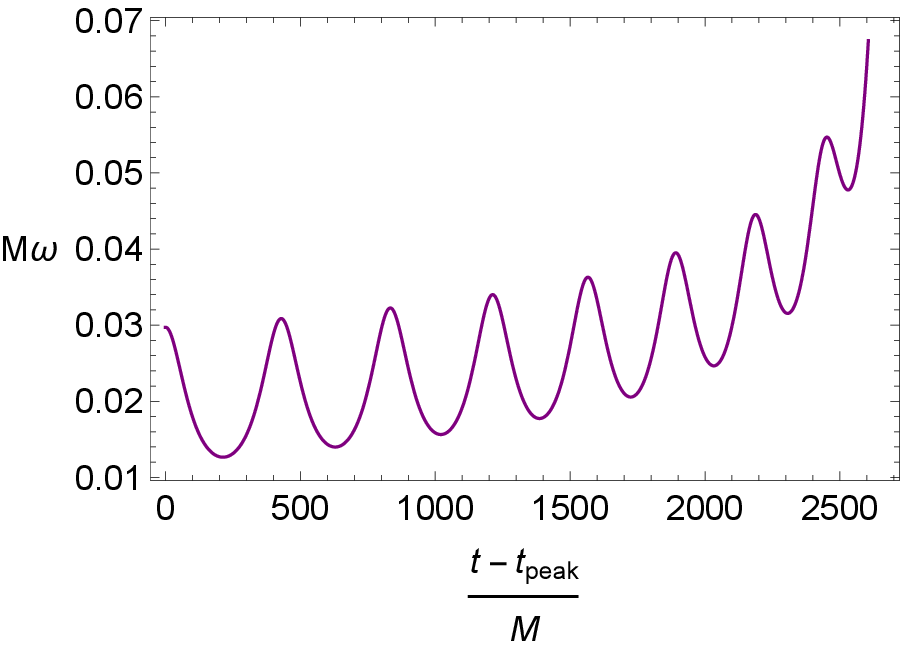}\\
		\includegraphics[scale=0.48]{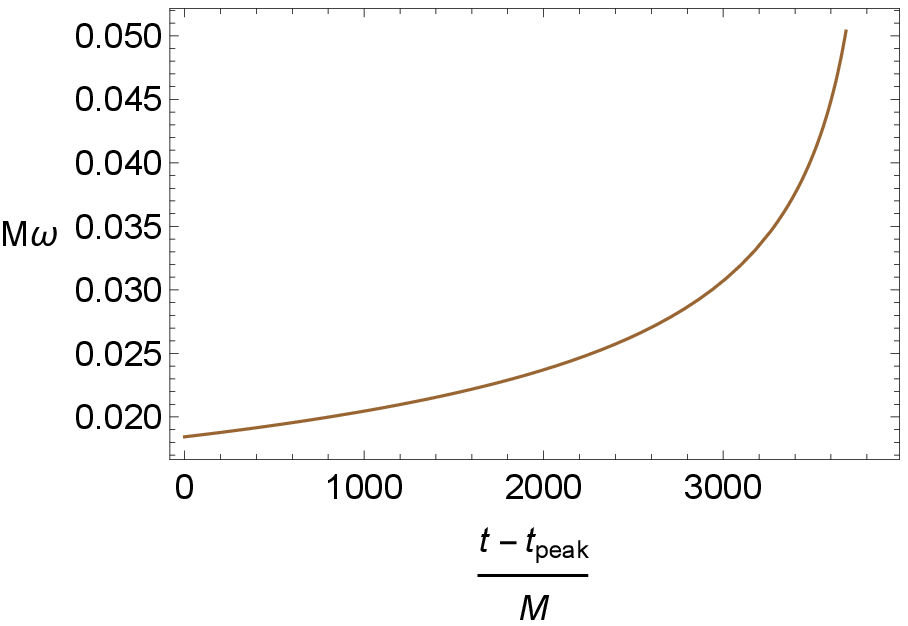}
		\includegraphics[scale=0.48]{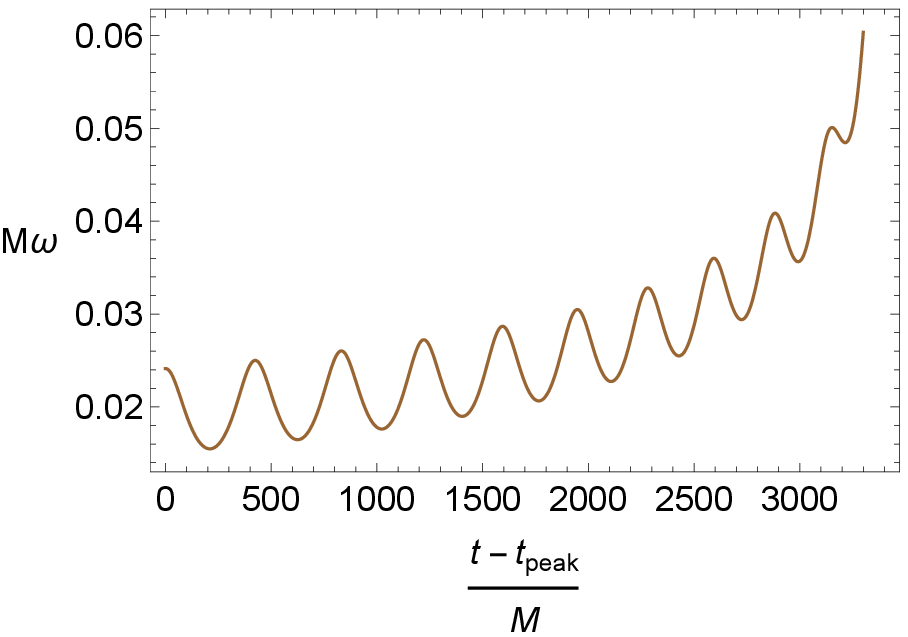}
		\includegraphics[scale=0.48]{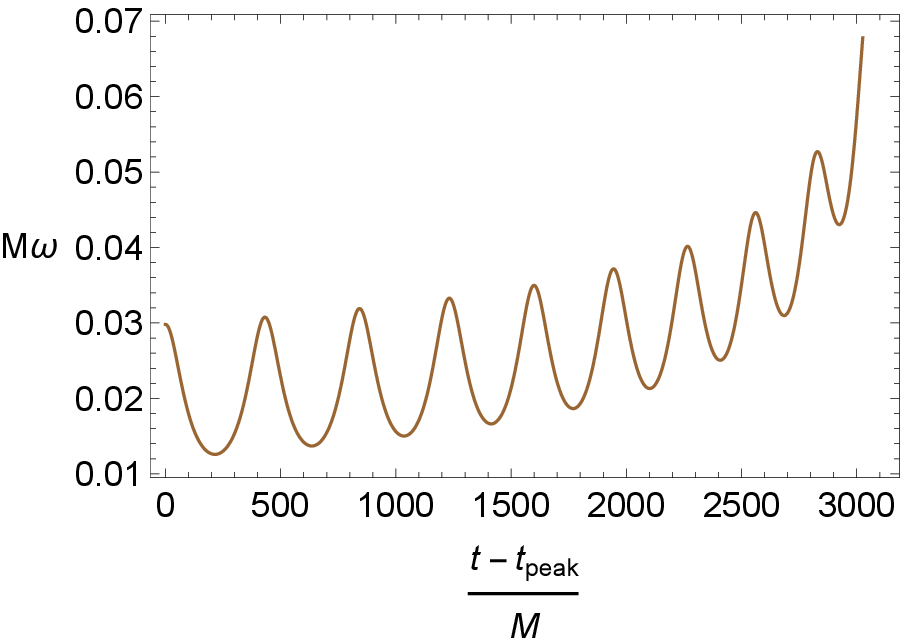}
		\caption{ Gravitational wave frequency as a function of time for the PN \textit{x model} plotted for different $ q $ and $ e $. Oscillatory behavior reduces with eccentricity. The colors follow the schema outlined in Figure \ref{fh}.}\label{fmw} 
	\end{figure}

	For a quasicircular situation, the frequency would vary monotonically, but the real data GW frequency in time revealed the presence of oscillations due to eccentricity. The entire time evolution is shown in Figure \ref{fmw}, for different $ q $. 
	
	The radial orbital period $ P $ conformed to the period of the oscillations in amplitude and instantaneous frequency. The amplitude of the oscillations was associated with the eccentricity $ e $. It increased with the eccentricity and the mean anomaly, $  l $.

\begin{figure}\centering
		\includegraphics[scale=0.48]{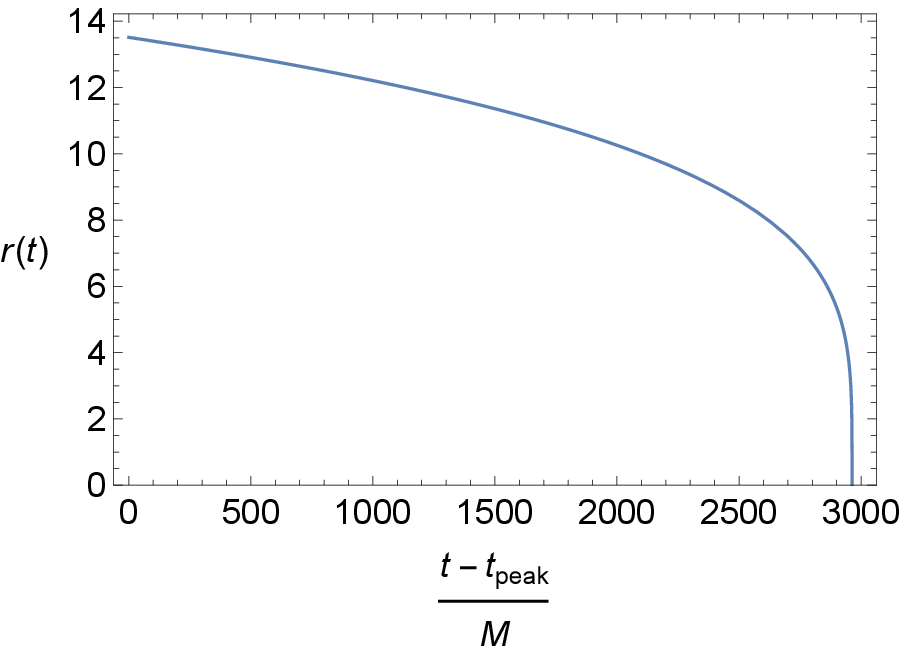}
		\includegraphics[scale=0.48]{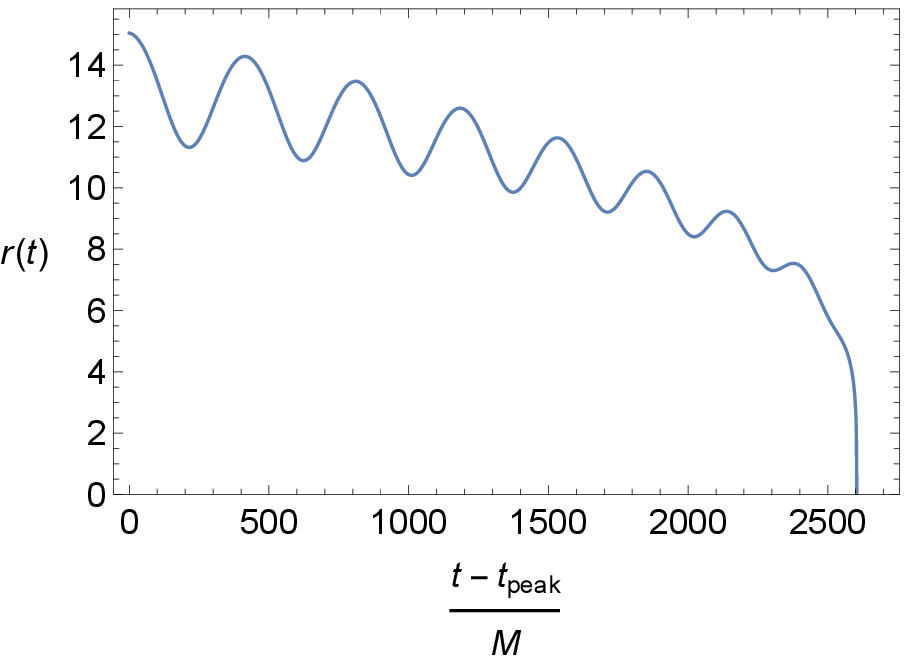}
		\includegraphics[scale=0.48]{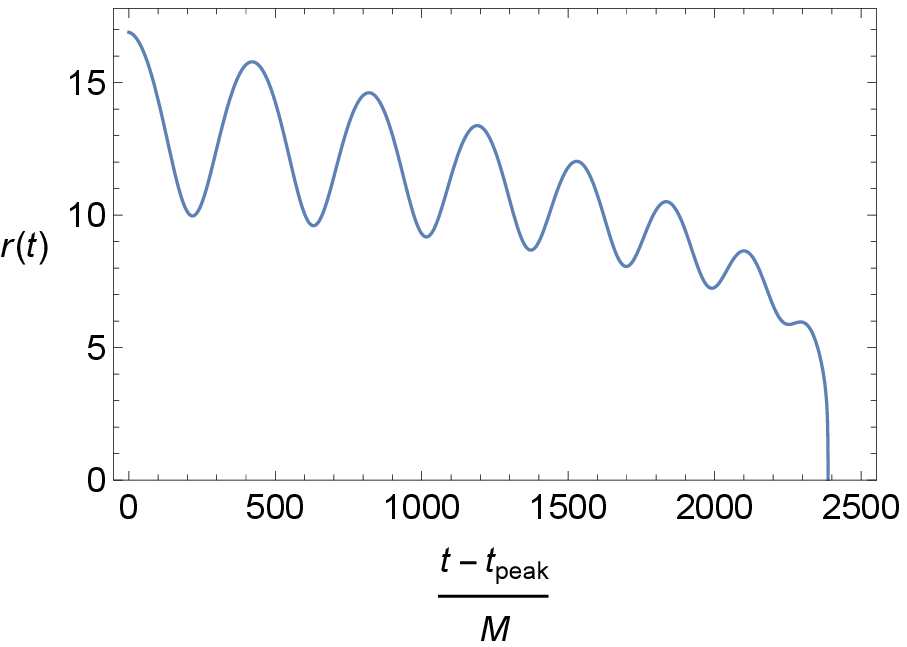}\\
		\includegraphics[scale=0.48]{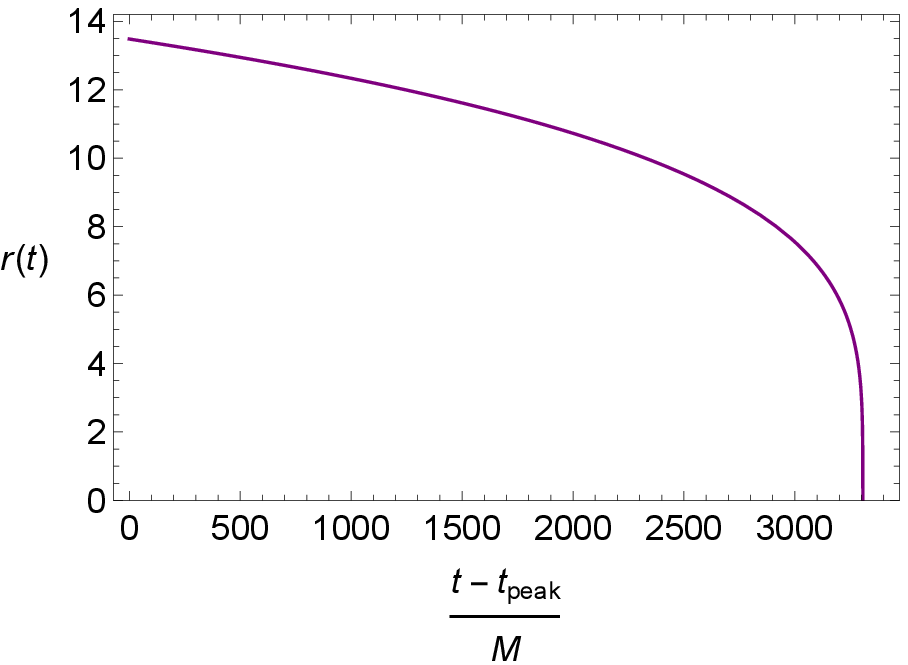}
		\includegraphics[scale=0.48]{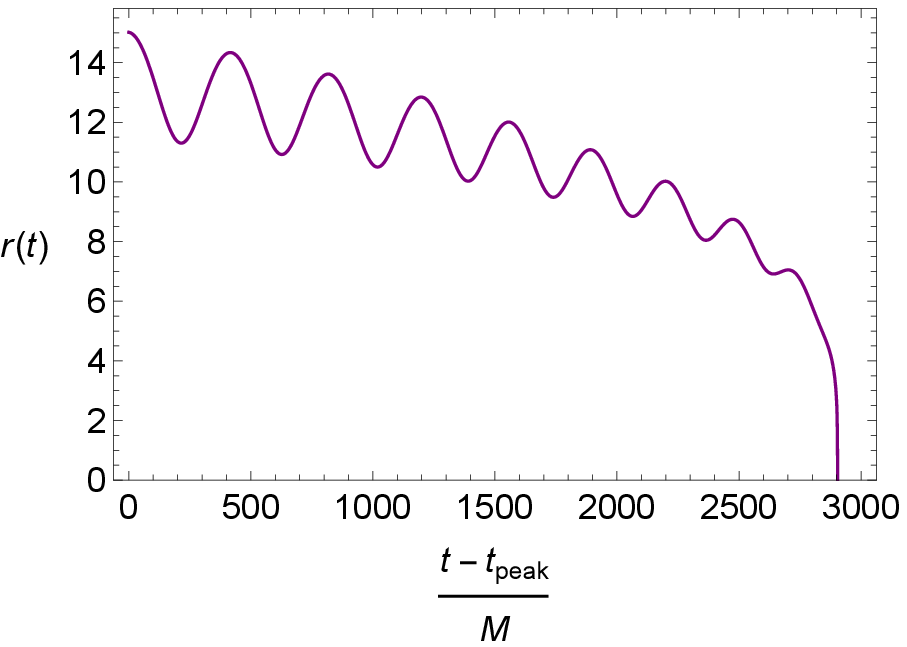}
		\includegraphics[scale=0.48]{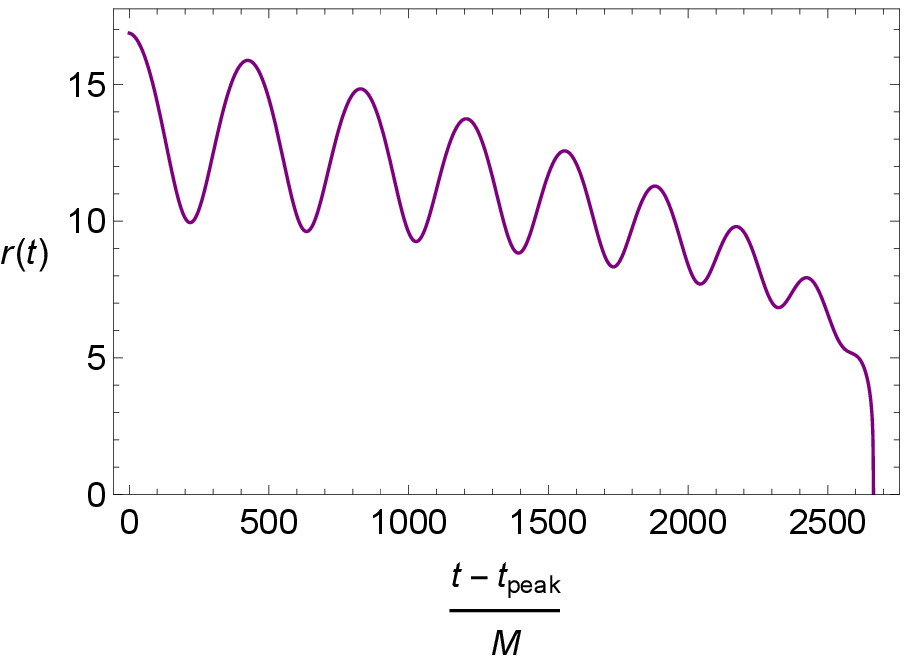}\\
		\includegraphics[scale=0.48]{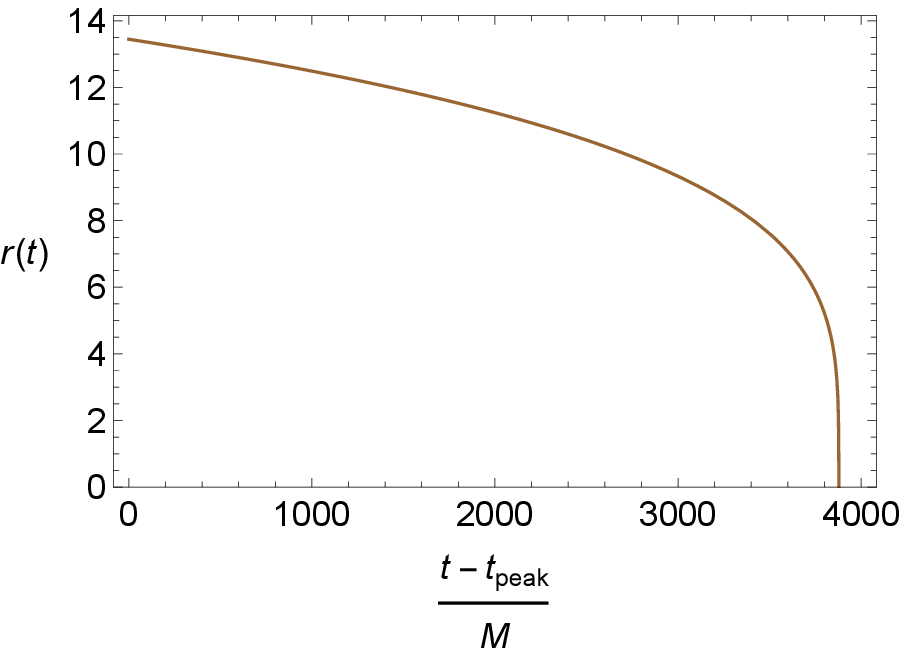}
		\includegraphics[scale=0.48]{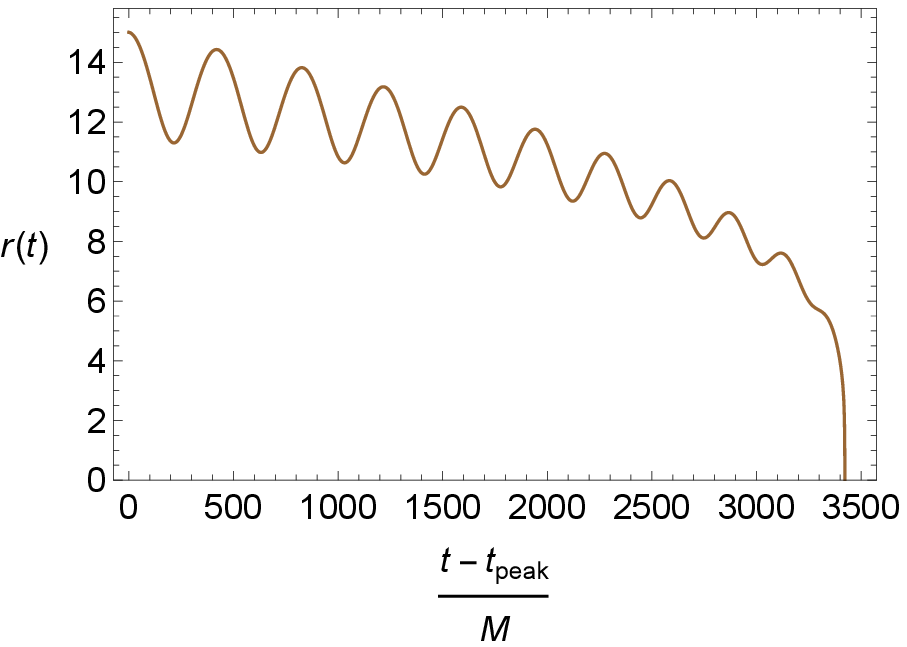}
		\includegraphics[scale=0.48]{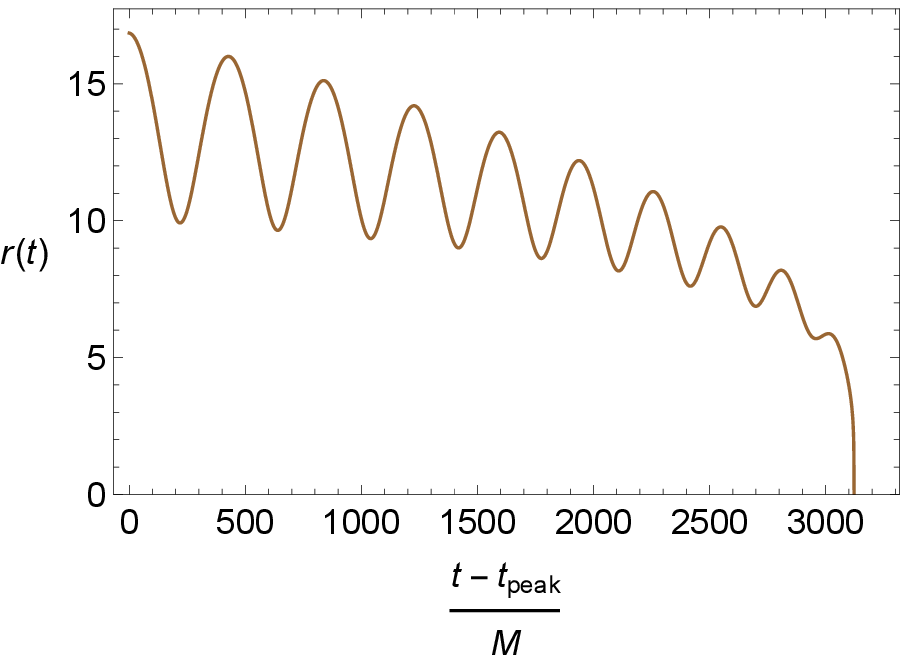}\\
		\caption{The variation in the orbital radius of the $ l = 2, m = 2 $ mode of the gravitational wave strain with $ (t-t_{peak})/M $. The simulation shows the oscillatory nature of $ r $ with the increases in $ e $. The colors follow the schema outlined in Figure \ref{fh}.}\label{fr}
	\end{figure}

	As the cycle increased, the orbital radius decreased, consistently initially; then, in the last few laps before merging, it decreased rapidly, as shown in Figure \ref{fr}. However, there was no oscillation for $ e \rightarrow 0 $. For non-null values of the eccentricity, the oscillation decreased, as the system approached  merging. The amplitude of the oscillation was higher for higher eccentricities, and the frequency of oscillation increased for higher mass ratios.

	\vspace{-8pt}
	
	\vspace{-8pt}

	\section{Conclusions}\label{sec4}
	
	In this paper, we presented a frequency related model with the variable $ x $. We studied extensively the equal and unequal-mass inspiral evolution and the characteristic behavior of the PN model, with a parallel comparison. A technique for the numerical evolution of BBH systems with minimal eccentricity  initial-data parameters was studied, similar to Ref \cite{Husa}. 
	
\begin{figure}[htp!]
		\includegraphics[scale=0.49]{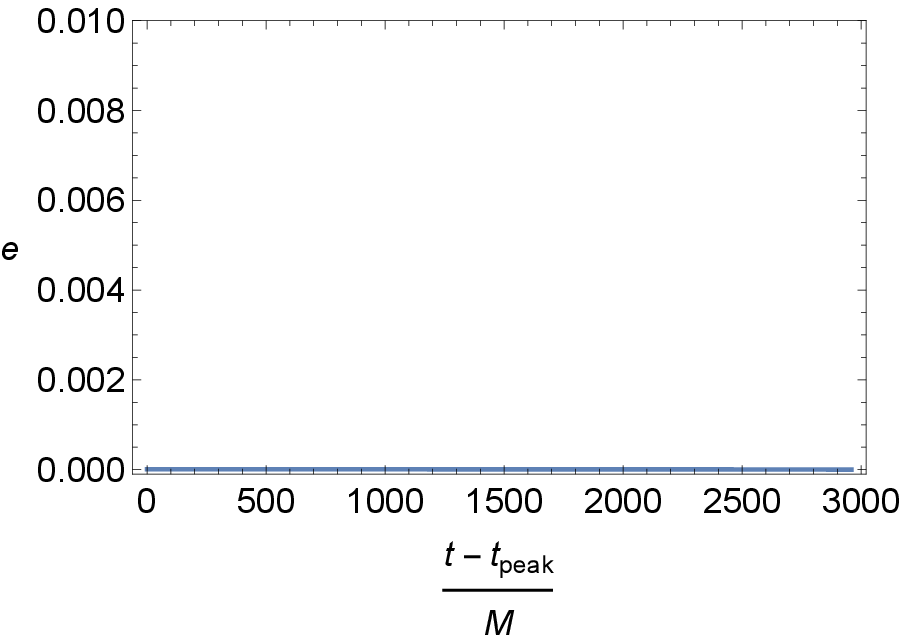}
		\includegraphics[scale=0.48]{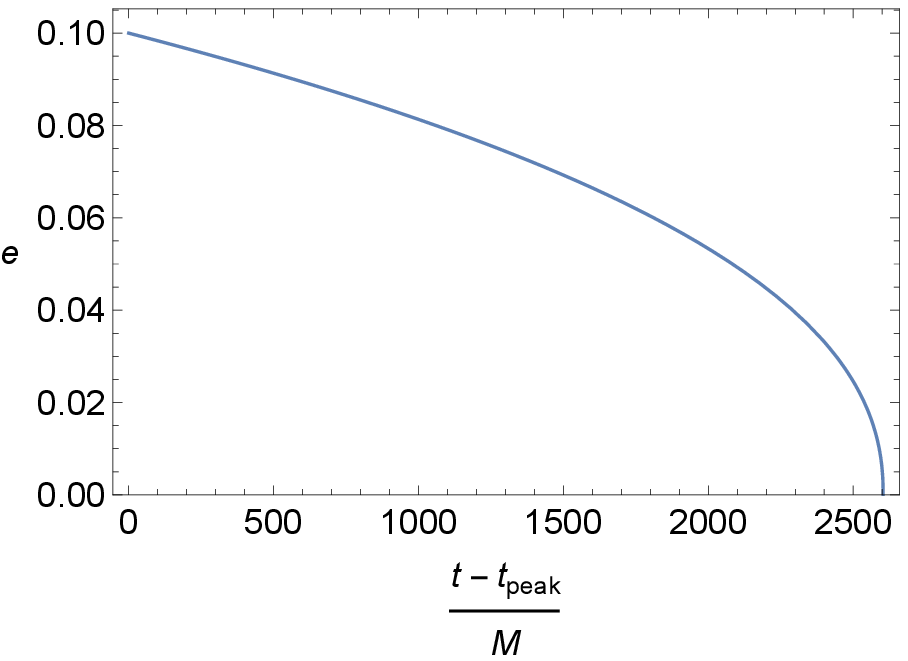}
		\includegraphics[scale=0.48]{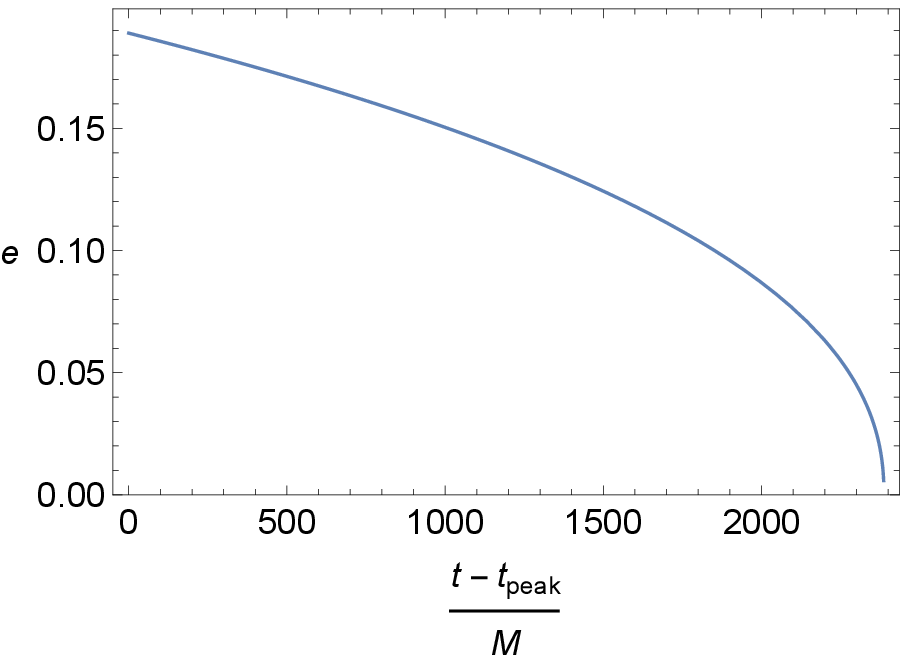}\\
		\includegraphics[scale=0.48]{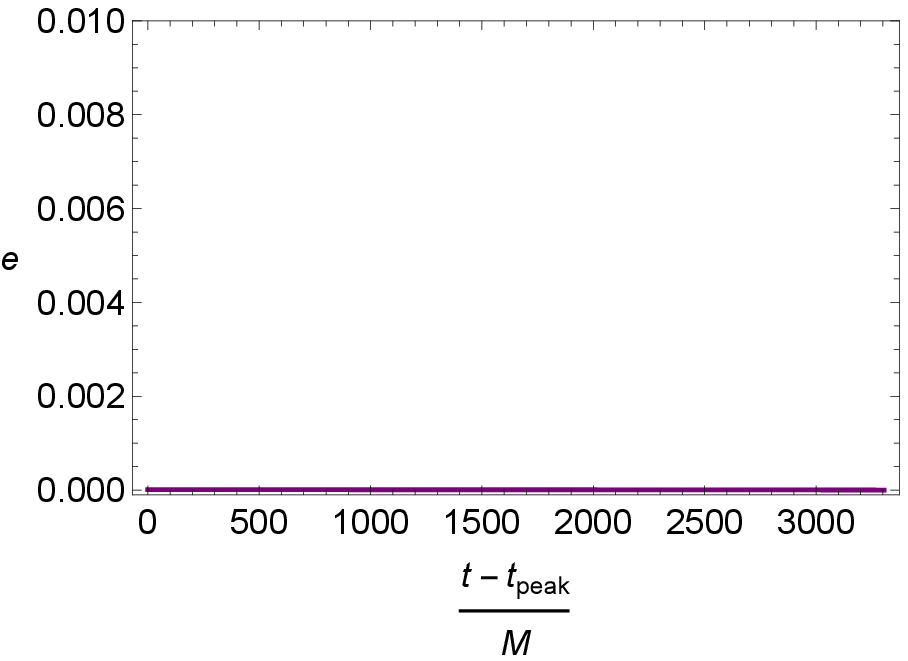}
		\includegraphics[scale=0.48]{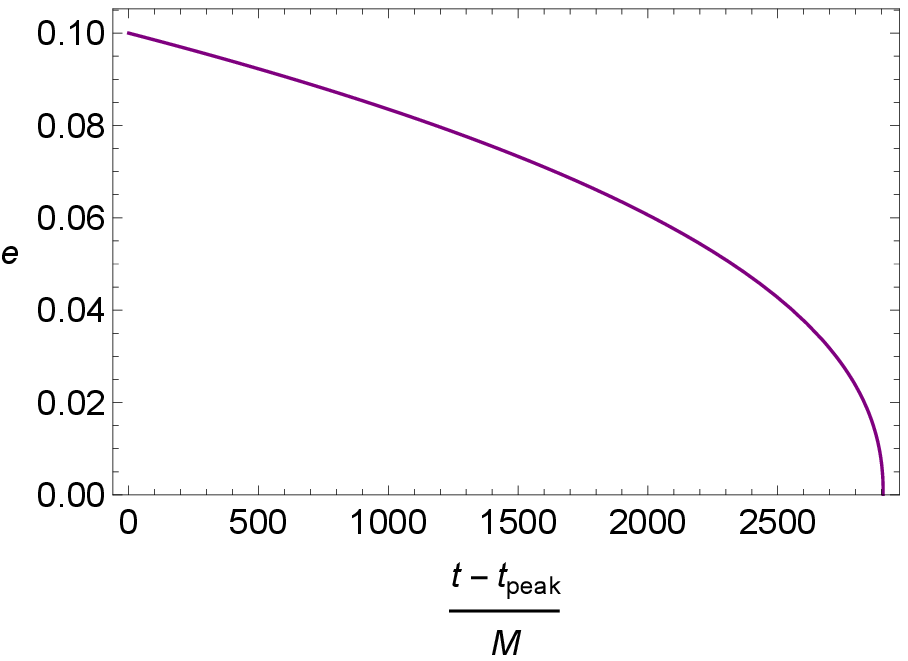}
		\includegraphics[scale=0.48]{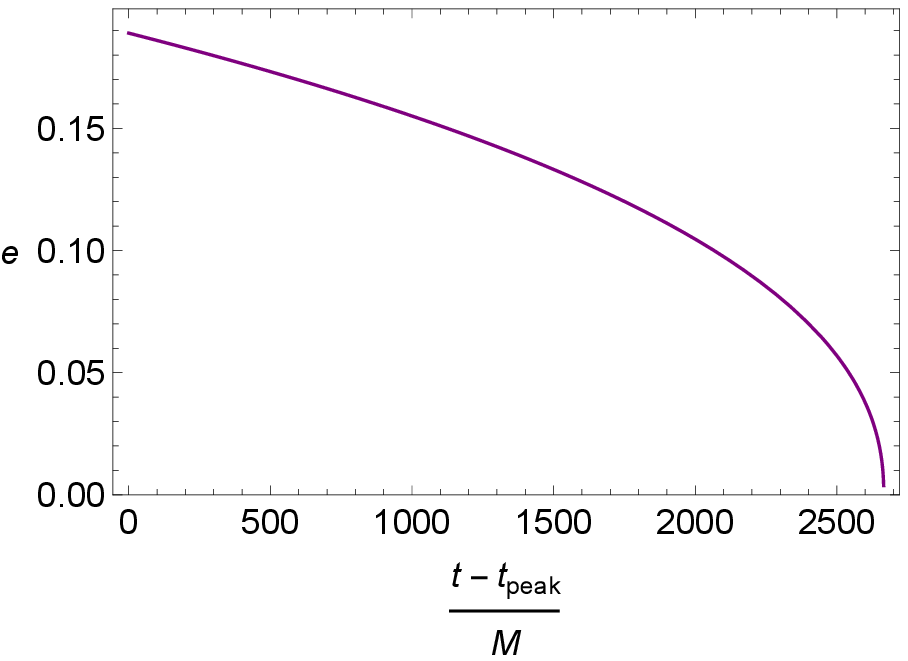}\\
		\includegraphics[scale=0.48]{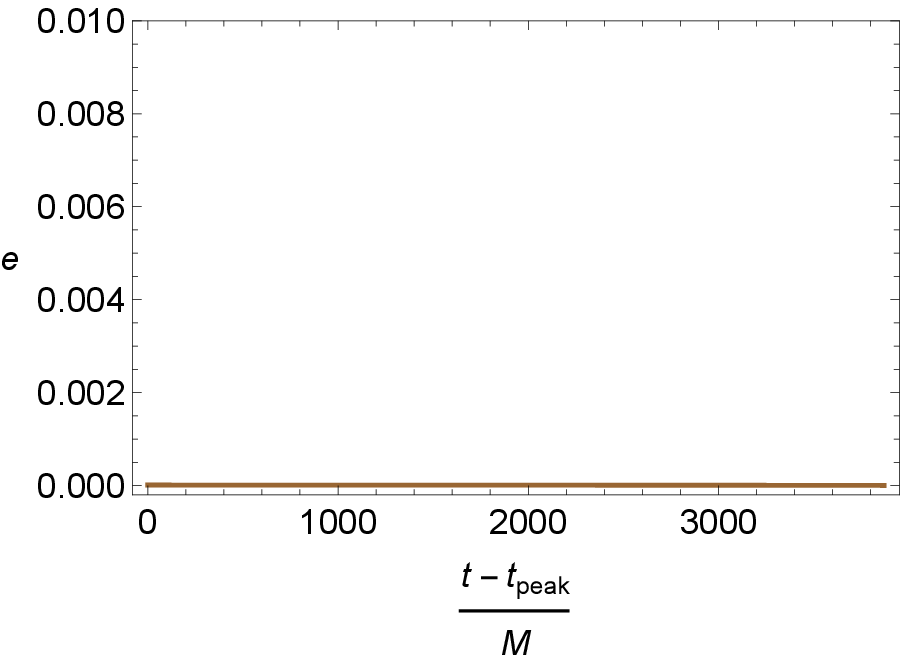}
		\includegraphics[scale=0.48]{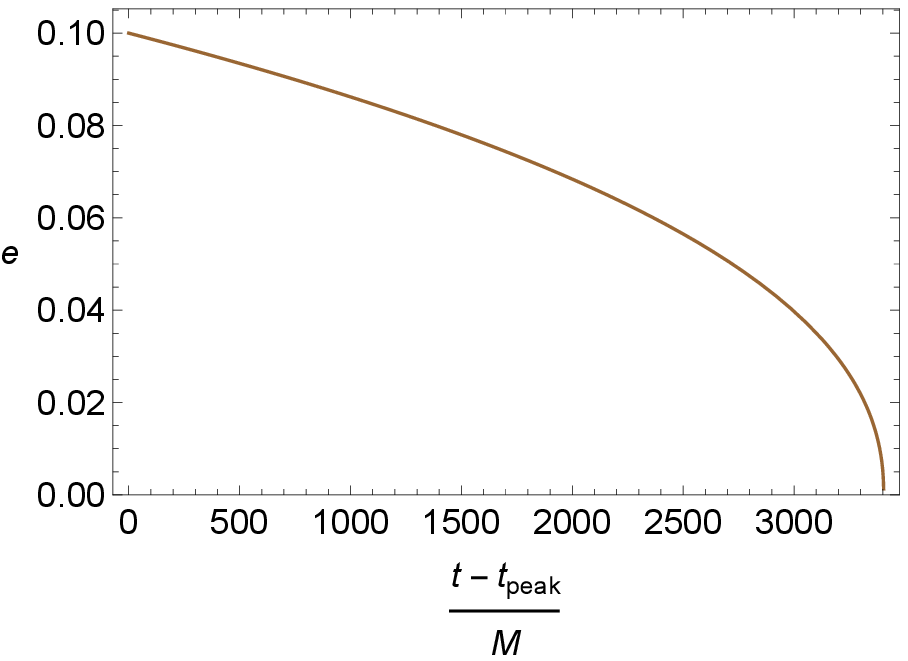}
		\includegraphics[scale=0.48]{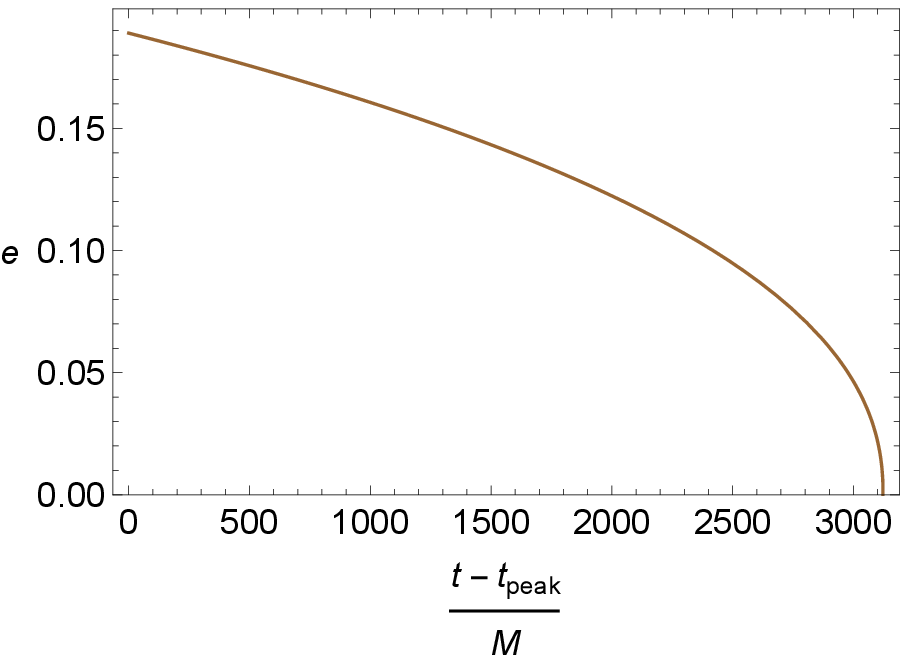}
		\caption{Time evolution of the $ e $ for different $ q $ . As $ q $ increases with $ e $, the time of the merger decreased significantly. The colors follow the schema outlined in Figure \ref{fh}.\label{fe}}
	\end{figure}

	For the BBH system, the eccentricity evolves with time. Since our model primarily depends on eccentricity, it was necessary to study the evolution of the eccentric character of the system. Figure \ref{fe} represents the entire scenario of the eccentricity. It was expected that in a binary system, before merging, the system would exhibit a circular motion. The binary system's well-defined non-zero initial eccentricity fell to zero at the final stage. 
	For such a system, the eccentricity decreased consistently. In the last $ 4\sim7 $ cycles before the merger, it rapidly fell to zero. 
	The BBH system with the initial zero eccentricity retained its null value throughout the evolution

	In the PN background, for the $ l=2, m=2 $ mode, the nature of the waveform, frequency, and phase were of a similar character. Based on the binary mass ratio, the variations shifted. These methods used dynamical entities, including some conjunction of the (2, 2) mode's amplitude or frequency, as the eccentricity evolved over time \cite{Peters}. Furthermore, the promising behavior of the frequency related variable $ x = (M\omega)^{2/3} $ motivated us to study the comparative behavior for different mass ratios. In the Newtonian limit, the approach minimized the eccentricity substantially. Eccentricity plays a key role in the model. The approximate eccentricity values of our system were the same as in Ref. \cite{Hinder1,Islam,url}. The value of the mean anomaly parametrized the waveform for $ t = t_{peak} $. The 2D parameter space for the model was as follows:
	$ \textrm{~eccentricity}:~e_{ref} \in [0,0.2]$ and $\textrm{~anomaly}:~ l_{ref} \in [-\pi, \pi] $. We considered the earliest conceivable fitting interval and sized it to roughly correlate to the preliminary orbital period, 433 M, in order to have a unique set of PN variables. This model was suboptimal, as it only had a 2 PN radiation reaction. Our \textit{x variable PN model} was very similar to the TaylorT4 model for $ e \rightarrow 0 $. Thus, same conclusions were also drawn here.
	
	Describing the nonspinning binaries was the central focus here. The inclusion of the effects of spin and higher modes would make this application more realistic with GW data. Finally, we can  assert that ``quasicircular'' waveforms are sufficient to detect GW. The minimum eccentricity was required to enable the most accurate fit to the PN inspiral waveforms.
	
	\vspace{6pt}
	

	
	
	\section*{Acknowledgments}The work of S.R.C. was supported by Southern Federal University (SFedU) (grant no. P-VnGr/21-05-IF). This research was supported in part by the International Centre for Theoretical Sciences (ICTS) for the online program---ICTS Summer School on Gravitational-Wave Astronomy (code: ICTS/gws2021/7). The research of M.K. was  supported by Southern Federal University, 2020 Project VnGr/2020-03-IF. The authors are also thankful to the anonymous referees for many suggestions, which have improved the manuscript.

	\section{\label{appendix}}	
	The fluxes of $ E $ and $ L $ are estimated by time average across a radial period P in the adiabatic estimation and are used to predict the rate change of parameters $ x $ and $ e $.	Such fluxes are used to 2 PN order in this application. 
	
	The separation between the masses	is 
	\begin{equation}
	r/M = (1-e \cos u)x^{-1}+ r_{1PN}+ r_{2PN}x + \mathcal{O}(x^2).
	\end{equation}
	
	The Kepler's equation in the PN form is
	\begin{equation}
	l = u- e \sin u + l_{2PN}x^2 + \mathcal{O}(x^3).
	\end{equation}
	
	The relative angular velocity is 
	\begin{equation}
	M \dot{\phi} = \frac{\sqrt{1 -e^2}}{(1-e \cos u)^2}x^{3/2} +\dot{\phi}_{1PN}x^{5/2} +\dot{\phi}_{2PN}x^{7/2}+ \mathcal{O}(x^{9/2}).
	\end{equation}
	
	The mean motion is explicitly expressed as 
	\begin{equation}
	M\dot{l} =Mn = x^{3/2} + n_{1PN}x^{5/2} +n_{2PN}x^{7/2} + \mathcal{O}(x^{9/2}).
	\end{equation}
	
	Here, $ r_{1PN},~ r_{2PN},~ l_{2PN},~ \dot{\phi}_{1PN}, \textrm{~and~} \dot{\phi}_{2PN} $ are functions of $ u $ and $ e $; whereas, $n_{1PN} $ and $n_{2PN}$ are only functions of $ e $.
	
	The expressions of the equations in the PN order are written in terms of the functions $k_E $ and ${k_J}$ in \cite{Gopa} and are also provided by K\"{o}nigsdo\"{o}rffer and Gopakumar \cite{Gopa1} in terms of $ n $ and $ e $ upto 3.5 PN, though we restricted our analysis in this study to 2PN.

		\section*{References}        

		

		
		

\end{document}